\documentclass{mem}
\usepackage{natbib}\usepackage{txfonts}\usepackage{balance}
\usepackage{graphicx}
\idline{75}{282}
\begin{document}
\def\teff{$T\rm_{eff }$}
\def\kms{$\mathrm {km s}^{-1}$}

\title{
Gamma-ray bursts as tracers of star-formation rate and metallicity evolution with THESEUS 
}

   \subtitle{}

\author{
S.D. \,Vergani\inst{1} 
          }

\institute{
GEPI, Observatoire de Paris, PSL Université, CNRS,  5 Place Jules Janssen, 92190 Meudon, France
%
\email{susanna.vergani@obspm.fr}
}

\authorrunning{Vergani }

\titlerunning{GRBs as tracers of SFR and metallicity evolution}

\abstract{
The class of long gamma-ray bursts (LGRBs) is associated with the collapse of the most massive stars, making them a tool to investigate
star formation in the early Universe.
Furthermore, thanks to their exceptional brightness, LGRB afterglows can be used as extra-galactic background sources
capable of unveiling the properties of the cold/warm gas of their hosting galaxies, up to the highest redshifts. Therefore LGRBs allow, uniquely, the combination of the information on different phases of the gas, through the absorption lines present in the afterglow spectra and the emission (continuum and lines) properties obtained from host galaxy photometry and spectroscopy, once the afterglow disappeared.

To date, the main problem to carry out this kind of studies at very high redshift is the poor number of GRBs detected at $z>5$.
The {\it THESEUS} mission will hugely increase such number, as it is expected to detect $>100$ GRBs at $z>5$. Therefore, {\it THESEUS} will bring a unique and fundamental contribution to the understanding of the chemical enrichment of the universe and of the evolution of galaxies and star formation up to the highest redshift.

}
\maketitle{}

\section{GRBs as tracers of star formation}

Long-duration GRBs (LGRBs) are produced by massive stars, and so
track star formation, and in particular the populations of
UV-bright stars responsible for the bulk of ionizing radiation
production.

Indeed, different studies pointed out the possibility to use LGRBs as tracers of star-formation activity up to the
highest redshifts (e.g.: \citealt{Robertson2012,Kistler2013}). The advantage of using LGRBs instead of galaxies is that GRB are detected through their gamma-ray emission and afterglows which are bright also at high redshift. 
On the other hand, most of past works suffer of the fact that only 30\% of GRBs has a measured redshift, and of the lack of statistics at high redshift.

\begin{figure*}[h!]
\includegraphics[width=7.3 cm]{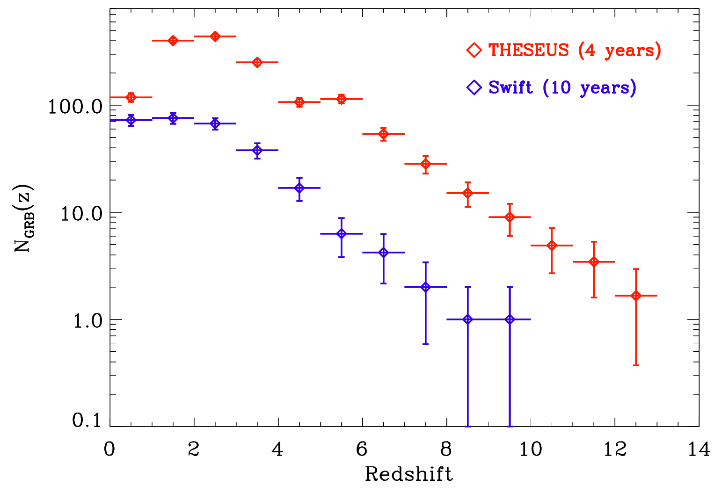}
\includegraphics[width=6.0 cm]{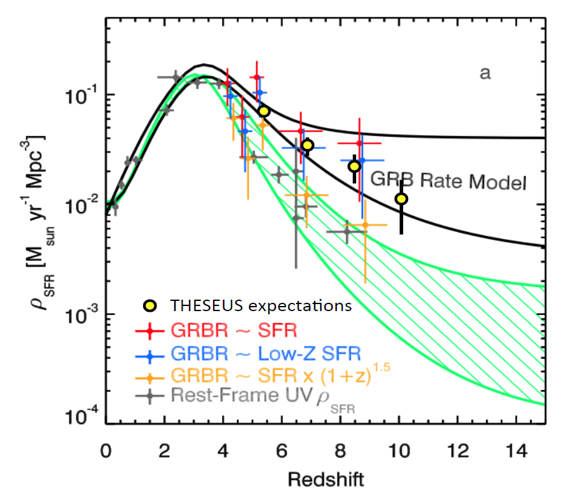}
\caption{\footnotesize
{\it Left}: Expected detection rate of GRBs, and uncertainty, of {\it THESEUS} compared to {\it Swift}.
The improvement of the number of very high-redshift GRBs reflects in a much smaller error on the rates and therefore 
on the SFR density in function of redshift as shown in the right panel. {\it Right}: SFR density evolution, 
adapted from \cite{Robertson2012}, showing the simulation of the information on the SFR density evolution 
that could be obtained thanks to the GRB detected by the {\it THESEUS} mission. {\it Credits}: {\it THESEUS} 
collaboration. 
}
\label{SFR}
\end{figure*}

In order to use LGRB as star-formation tracers and LGRB host galaxies as a representative population
of star-forming galaxies, we need to understand the link between the LGRB rate and star formation rate,
in the following referred to as LGRB efficiency.
Theoretically, there are various reasons to expect a dependence of the LGRB efficiency on metallicity. Indeed, most theoretical models predict low-metallicity for the GRB progenitor star as a necessary condition to eliminate the progenitor star hydrogen envelope (which
would smother the jet, and produce emission lines in the GRB-associated supernova that are not observed) without
spinning down the angular momentum that the central engine needs to launch the jet.

Using the {\it Swift/BAT6} \citep{Salvaterra2012} and SHOALS \citep{Perley2016b} complete samples of LGRB host galaxies, respectively, \cite{Vergani2015,Vergani2017}, \cite{Japelj2016}, and \cite{Perley2016} showed that LGRBs prefer sub-solar metallicity host galaxies. This implies that, if metallicity is the only factor affecting the LGRB efficiency, LGRBs can be used to trace star formation rate (SFR) at high-redshift (see also \citealt{Greiner2015}) as at that epoch the bulk of star-forming galaxies fulfil the condition of low-metallicity required by LGRBs.

During four years of operations {\it THESEUS} will detect more than 100 LGRBs at $z>5$, $\sim10$ of which being at $z>10$. The substantial improvement in the GRB detected at $z>7-8$ will reduce the uncertainties on the rates and bring to tighter constraints on the SFR density in function of redshift (see Fig.\ref{SFR}).

\section{GRBs as tracers of metallicity evolution}

\begin{figure*}[h!]
\centering
\includegraphics[width=7.2 cm]{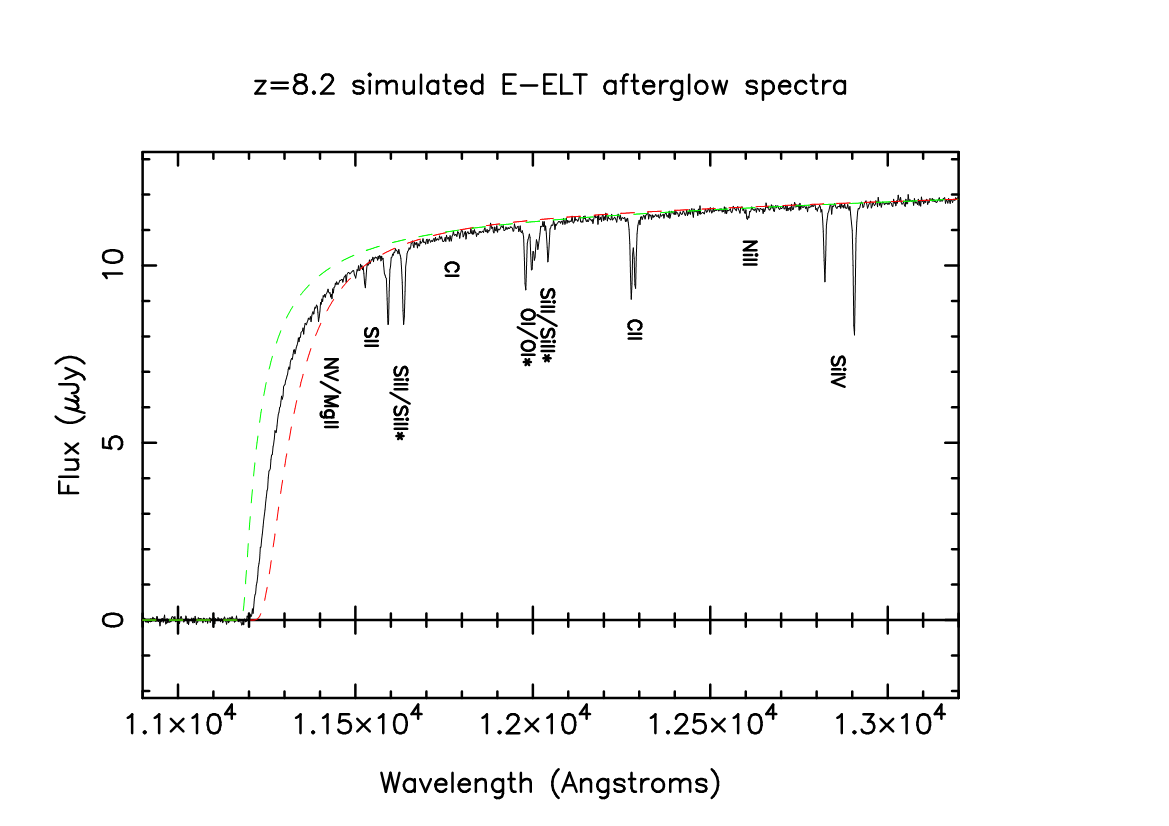}
\includegraphics[width=6.2 cm]{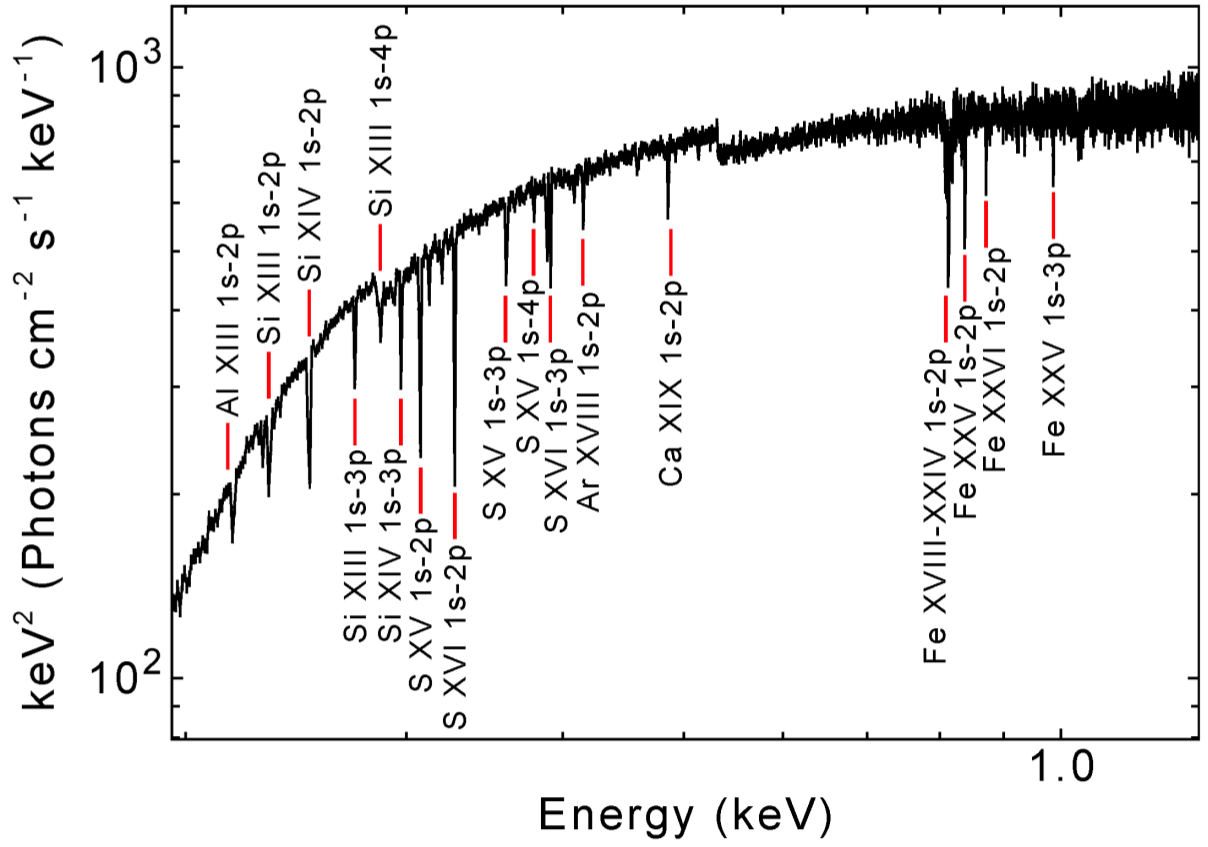}
\includegraphics[width=8.0 cm]{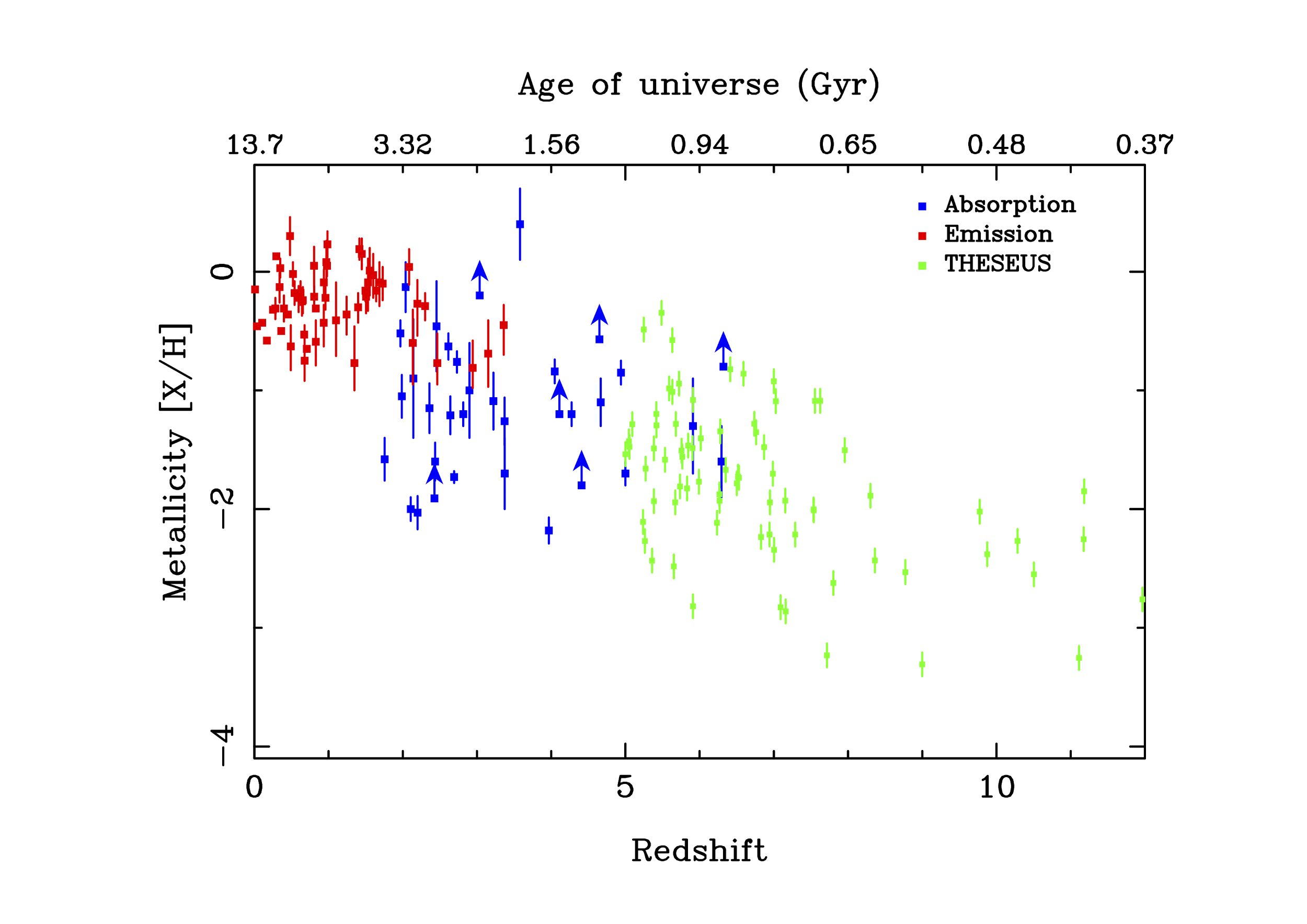}
\caption{\footnotesize 
{\it Top left}: Simulated ELT (European Extremely Large Telescope) spectrum of a GRB at $z=8.2$ as discovered
by {\it THESEUS} (providing also the arcsec localization and redshift measurement). The S/N provides exquisite abundance determinations
from metal absorption lines, while fitting the Ly-$\alpha$\,damping wing simultaneously
fixes the IGM neutral fraction and the host HI column
density, as illustrated by the two extreme models: a pure 100\% neutral
IGM (green) and best-fit host absorption with a fully ionized IGM
(red). From the {\it THESEUS White Paper} (arXiv:1710.04638).
{\it Top right}: Simulated ATHENA/X-IFU spectrum of a GRB at $z=7$ from \citet{jonker13} 
for which {\it THESEUS} could provide the detection, arcsec localization, and redshift measurement. 
As can be seen, the combination of the unique capabilities of THESEUS with the excellent sensitivity and spectroscopic 
performances of these future facilities (e.g., ATHENA/X-IFU and ELT) would allow an unprecedented study and comparison 
of the average composition of the host galaxy and of the circum-burst environment.
{\it Bottom}: Absorption-line based metallicities [M/H] as a function of redshift for GRB-DLAs (blue symbols) and emission line metallicities
for GRB hosts (red symbols; adapted from \citealt{Sparre2014}).
The green points represent the results of the simulation of the metallicity obtained using a sample of PopII GRB detected by 
{\it THESEUS} at $z>5$. The number of events corresponds to the minimum numbers of the expected GRB detected by {\it THESEUS}. 
{\it Credits}: {\it THESEUS} collaboration.
}
\label{met}
\end{figure*}

Thanks to their exceptional brightness, only marginally decreasing with redshift, GRB afterglows can be used as powerful extragalactic background sources. The afterglow spectrum reveals the signatures of the gas associated
with the GRB host, but also with the IGM and with other galaxies present along the GRB line of sight.
The spectroscopy of LGRB optical afterglows provide a unique tool to unveil the properties of the
ISM even in the faintest and most distant galaxies. The analysis of the absorption lines bring to a
direct measurement of the column densities of the different elements present in the interstellar medium (ISM; neutral hydrogen, metals, molecules...). The most
prominent absorption line seen in high-redshift GRB spectra is the Lyman-$\alpha$\, line of
neutral hydrogen (HI). From the fitting of the damped Lyman-$\alpha$\, (DLA) profile, it is possible to obtain
accurate measurement of the neutral hydrogen column density (NHI). The absorption lines of different metal species, 
both in their low- and high-ionization state, are also commonly detected. The ratio of the metal and NHI
column densities give a direct determination of the metallicity of the gas. The metallicity and abundance
measurements of the ISM of LGRB host galaxies can be used to constrain the chemical enrichment of
star-forming galaxies in the Universe (see Fig.\,\ref{met}).
Indeed, the association of LGRBs with massive stars also means they are located within the disks /
inner regions of typical star-forming galaxies. Differently from quasars that probe random lines of sight
through the foreground galaxies, LGRBs probe the gas associated with the star-forming
regions of their host galaxy.

{\it THESEUS} will identify a high number of GRBs at high redshift (more than 100 at $z>5$), providing also the photometric redshift 
determination for all of them. Therefore, it will be possible to select these very high-redshift GRBs so as to obtain high-quality 
afterglow spectroscopy from the ground-based facilities, identifying the ISM absorption features even at the highest redshifts 
(Fig.\,\ref{met}). Thanks to these observations, it will be possible to study the chemical evolution of star-forming 
galaxies, in principle even beyond $z=10$ (Fig.\,\ref{met}).

\section{Conclusions}
{\it THESEUS} will have a unique impact in the studies of the early universe and galaxy evolution, by hugely improving the GRB statistics at high redshift and identifying very-high redshift GRBs. Both factors are fundamental and will allow a high-impact use of GRBs as probes of the high-redshift Universe, determining the evolution of the star-formation rate density and metallicity up to the early Universe.

GRBs offer the unique opportunity to investigate systematically (and in principle at any redshift) both
the cold/warm gas in the ISM (using the absorption lines seen in the afterglow spectra) and the ionised gas (using the
emission lines of the host galaxy spectra). For high-redshift faint star-forming galaxies, the combination of the information obtained from both diagnostics will not be possible through 
{\it standard} galaxy studies, even with the {\it JWST}. 
Instead, LGRBs offer the possibility to reach this goal up to the highest redshift.
The ground- and space-based follow-up of the significant number of high-redshift GRBs detected by THESEUS, will provide a unique and detailed picture of the properties of the population of high-redshift galaxies.

\bibliographystyle{aa}
\bibliography{susy16}

\begin{thebibliography}{11}
\expandafter\ifx\csname natexlab\endcsname\relax\def\natexlab#1{#1}\fi

\bibitem[{{Greiner} {et~al.}(2015){Greiner}, {Fox}, {Schady}, {Kr{\"u}hler},
  {Trenti}, {Cikota}, {Bolmer}, {Elliott}, {Delvaux}, {Perna}, {Afonso},
  {Kann}, {Klose}, {Savaglio}, {Schmidl}, {Schweyer}, {Tanga}, \&
  {Varela}}]{Greiner2015}
{Greiner}, J., {Fox}, D.~B., {Schady}, P., {et~al.} 2015, \apj, 809, 76

\bibitem[{{Japelj} {et~al.}(2016){Japelj}, {Vergani}, {Salvaterra}, {D'Avanzo},
  {Mannucci}, {Fernandez-Soto}, {Boissier}, {Hunt}, {Atek},
  {Rodr{\'{\i}}guez-Mu{\~n}oz}, {Scodeggio}, {Cristiani}, {Le Floc'h},
  {Flores}, {Gallego}, {Ghirlanda}, {Gomboc}, {Hammer}, {Perley}, {Pescalli},
  {Petitjean}, {Puech}, {Rafelski}, \& {Tagliaferri}}]{Japelj2016}
{Japelj}, J., {Vergani}, S.~D., {Salvaterra}, R., {et~al.} 2016, \aap, 590,
  A129

\bibitem[{{Jonker} {et~al.}(2013){Jonker}, {O'Brien}, {Amati}, {Atteia},
  {Campana}, {Evans}, {Fender}, {Kouveliotou}, {Lodato}, {Osborne}, {Piro},
  {Rau}, {Tanvir}, \& {Willingale}}]{jonker13}
{Jonker}, P., {O'Brien}, P., {Amati}, L., {et~al.} 2013, arXiv/1306.2336

\bibitem[{{Kistler} {et~al.}(2013){Kistler}, {Yuksel}, \&
  {Hopkins}}]{Kistler2013}
{Kistler}, M.~D., {Yuksel}, H., \& {Hopkins}, A.~M. 2013, ArXiv:1305.1630

\bibitem[{{Perley} {et~al.}(2016{\natexlab{a}}){Perley}, {Kr{\"u}hler},
  {Schulze}, {de Ugarte Postigo}, {Hjorth}, {Berger}, {Cenko}, {Chary},
  {Cucchiara}, {Ellis}, {Fong}, {Fynbo}, {Gorosabel}, {Greiner}, {Jakobsson},
  {Kim}, {Laskar}, {Levan}, {Micha{\l}owski}, {Milvang-Jensen}, {Tanvir},
  {Th{\"o}ne}, \& {Wiersema}}]{Perley2016b}
{Perley}, D.~A., {Kr{\"u}hler}, T., {Schulze}, S., {et~al.} 2016{\natexlab{a}},
  \apj, 817, 7

\bibitem[{{Perley} {et~al.}(2016{\natexlab{b}}){Perley}, {Tanvir}, {Hjorth},
  {Laskar}, {Berger}, {Chary}, {de Ugarte Postigo}, {Fynbo}, {Kr{\"u}hler},
  {Levan}, {Micha{\l}owski}, \& {Schulze}}]{Perley2016}
{Perley}, D.~A., {Tanvir}, N.~R., {Hjorth}, J., {et~al.} 2016{\natexlab{b}},
  \apj, 817, 8

\bibitem[{{Robertson} \& {Ellis}(2012)}]{Robertson2012}
{Robertson}, B.~E. \& {Ellis}, R.~S. 2012, \apj, 744, 95

\bibitem[{{Salvaterra} {et~al.}(2012){Salvaterra}, {Campana}, {Vergani},
  {Covino}, {D'Avanzo}, {Fugazza}, {Ghirlanda}, {Ghisellini}, {Melandri},
  {Nava}, {Sbarufatti}, {Flores}, {Piranomonte}, \&
  {Tagliaferri}}]{Salvaterra2012}
{Salvaterra}, R., {Campana}, S., {Vergani}, S.~D., {et~al.} 2012, \apj, 749, 68

\bibitem[{{Sparre} {et~al.}(2014){Sparre}, {Hartoog}, {Kr{\"u}hler}, {Fynbo},
  {Watson}, {Wiersema}, {D'Elia}, {Zafar}, {Afonso}, {Covino}, {de Ugarte
  Postigo}, {Flores}, {Goldoni}, {Greiner}, {Hjorth}, {Jakobsson}, {Kaper},
  {Klose}, {Levan}, {Malesani}, {Milvang-Jensen}, {Nardini}, {Piranomonte},
  {Sollerman}, {S{\'a}nchez-Ram{\'{\i}}rez}, {Schulze}, {Tanvir}, {Vergani}, \&
  {Wijers}}]{Sparre2014}
{Sparre}, M., {Hartoog}, O.~E., {Kr{\"u}hler}, T., {et~al.} 2014, \apj, 785,
  150

\bibitem[{{Vergani} {et~al.}(2017){Vergani}, {Palmerio}, {Salvaterra},
  {Japelj}, {Mannucci}, {Perley}, {D'Avanzo}, {Kr{\"u}hler}, {Puech},
  {Boissier}, {Campana}, {Covino}, {Hunt}, {Petitjean}, \&
  {Tagliaferri}}]{Vergani2017}
{Vergani}, S.~D., {Palmerio}, J., {Salvaterra}, R., {et~al.} 2017, \aap, 599,
  A120

\bibitem[{{Vergani} {et~al.}(2015){Vergani}, {Salvaterra}, {Japelj}, {Le
  Floc'h}, {D'Avanzo}, {Fernandez-Soto}, {Kr{\"u}hler}, {Melandri}, {Boissier},
  {Covino}, {Puech}, {Greiner}, {Hunt}, {Perley}, {Petitjean}, {Vinci},
  {Hammer}, {Levan}, {Mannucci}, {Campana}, {Flores}, {Gomboc}, \&
  {Tagliaferri}}]{Vergani2015}
{Vergani}, S.~D., {Salvaterra}, R., {Japelj}, J., {et~al.} 2015, \aap, 581,
  A102

\end{thebibliography}

\end{document}